\documentclass[twocolumn,10pt]{article}

\usepackage{fullpage}
\usepackage{authblk}
\usepackage{amsmath}
\usepackage{graphicx}
\usepackage{hyperref}
\usepackage{dblfloatfix}
\usepackage[usenames,dvipsnames]{color}

\title{
Machine learning for crystal identification and discovery
}

\author[1,2]{Matthew Spellings}
\author[1,2]{Sharon C. Glotzer\thanks{sglotzer@umich.edu}}
\affil[1]{Department of Chemical Engineering, University of Michigan, Ann Arbor, MI 48109}
\affil[2]{Biointerfaces Institute, University of Michigan, Ann Arbor, MI 48109}

\begin{document}
\maketitle

\abstract{
As computers get faster, researchers --- not hardware or algorithms --- become the bottleneck in scientific discovery. Computational study of colloidal self-assembly is one area that is keenly affected: even after computers generate massive amounts of raw data, performing an exhaustive search to determine what (if any) ordered structures occur in a large parameter space of many simulations can be excruciating. We demonstrate how machine learning can be applied to discover interesting areas of parameter space in colloidal self assembly. We create numerical fingerprints --- inspired by bond orientational order diagrams --- of structures found in self-assembly studies and use these descriptors to both find interesting regions in a phase diagram and identify characteristic local environments in simulations in an automated manner for simple and complex crystal structures. Utilizing these methods allows analysis methods to keep up with the data generation ability of modern high-throughput computing environments.
}

\section*{Introduction}

\begin{figure*}[h!tb]
\begin{center}
\includegraphics[width=\textwidth]{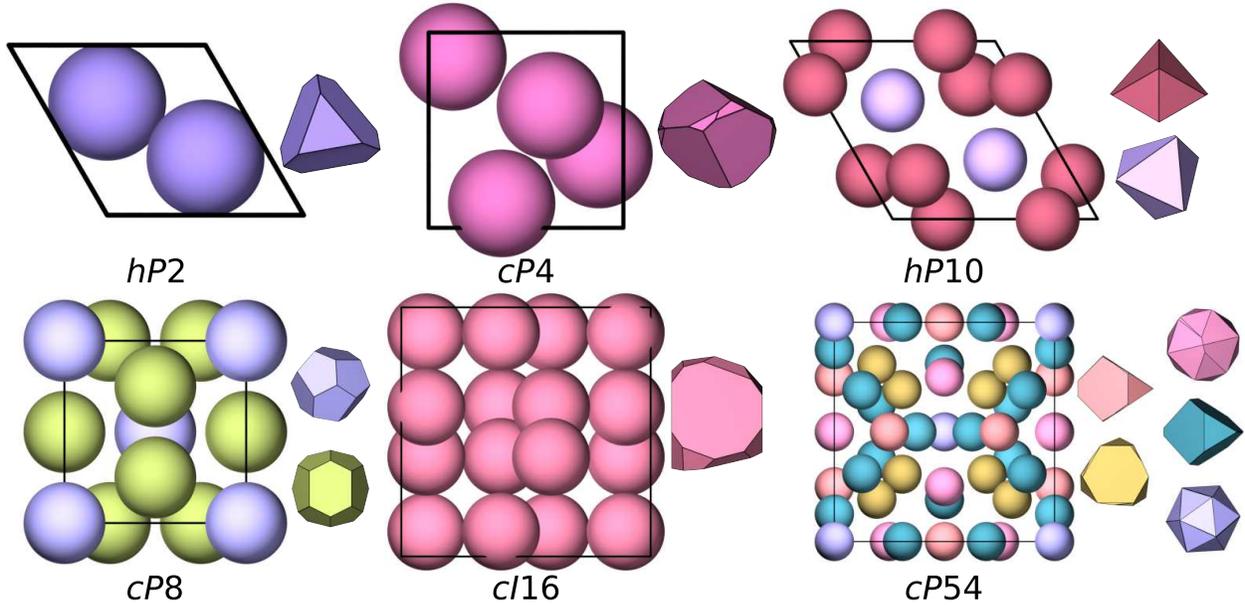}
\caption{A subset of the simple and complex structures that self-assemble from an isotropic pair potential described in~\cite{engel2015}. For each structure, Pearson symbols and particle configurations in crystal unit cells are shown on the left, and Voronoi polyhedra corresponding to representative nearest-neighbor local environments are shown on the right.}
\label{fig:complex_structures}
\end{center}
\end{figure*}

In the process of engineering the self-assembly behavior of colloidal- and nanoscale particles, scientists leave enormous amounts of configurational data in their wake. Experimentally, crystal structures with tunable properties can be created through anisotropic colloidal building blocks~\cite{henzie2012}, DNA-coated nanoparticles~\cite{shevchenko2006,macfarlane2011,mirkinProgrammable}, or a host of other interactions\cite{li2016}. In computational studies of colloidal matter, various simple, as well as complex, phases can be formed through systematic modification of entropic or enthalpic interparticle interactions~\cite{hynninen2006,glaser2007,batten2011,costacampos2012,damasceno2012,engel2015}. In an inverse design scenario, digital alchemy~\cite{vananders2015} allows us to quickly find building block attributes and thermodynamic conditions that optimize self assembly into a given target structure. For exploratory studies, however, the design process is more difficult. After the computationally expensive undertaking of performing simulations, each dataset must also be analyzed --- a procedure that is often manual, repetitive, and labor-intensive in the case of crystal structure identification. This analysis difficulty is partly due to the wide variety of crystal structures that can be found in self-assembling systems, as shown in Figure~\ref{fig:complex_structures}. As advances in hardware and software conspire to decrease the cost of parallel computation, it will only become more imperative that researchers utilize automated, high-performance analysis methods to investigate the data generated from their high-throughput simulation codes.

Automated analysis of data from two-dimensional systems has been successfully performed using variations on the \emph{n}-atic order parameter(s) $\psi_n$, defined for each particle as\cite{bernard2011,engel2013}

\[\psi_n = \frac{1}{n} \sum\limits_{j} e^ { in\theta_{ij} }. \]

\noindent where the sum is over particle $i$'s neighbors and $\theta_{ij}$ is the angle of the bond between particle $i$ and particle $j$. $\psi_n$ can identify tetratic and hexatic behavior in hard squares~\cite{wojciechowski2004}, rectangles~\cite{donev2006}, and disks\cite{bernard2011,engel2013}, as well as hexagonal order in systems of active disks\cite{redner2013}. In general, this order parameter works well for detecting local $n$-fold bond orientational ordering in two-dimensional systems. In three spatial dimensions, however, structural order can be more complex, and detecting it more challenging. The Steinhardt order parameter(s) $Q_n$~\cite{steinhardt1983} are natural three-dimensional analogues to the $n$-atic order parameters and they have been used to analyze many systems assembling relatively simple structures\cite{steinhardt1983,duijneveldt1992,yan2005}, but they have some shortcomings. Even for some of the simplest, most common structures we find in self-assembly, Steinhardt order parameters $Q_n$ (and the related family of order parameters $W_n$, also derived from combinations of neighbor-bond spherical harmonics) often poorly distinguish between distinct structures, and can be distributed differently for the same structure formed by dissimilar pairwise interactions~\cite{lechner2008}. Usually they must be carefully tuned by hand to optimize their specificity for each system they will be used to identify\cite{steinhardt1983,wolde1996,chau1998}. Ideally, the order parameters we use would be more robust and less biased if driven by the data we are interested in rather than arbitrary choices of symmetries to search for and threshold values for the chosen parameters.

Another problem that hinders automatic structure analysis is that we typically do not know which structures are present in a dataset before analyzing it exhaustively. This can be problematic even for simple systems. For example, hard particles --- which have some of the simplest interactions to define --- are known to self-assemble into a great diversity of complex structures, including quasicrystals and crystals with many-atom repeat units~\cite{amirTetrahedra,damasceno2012}. Creating and tuning high-specificity order parameters by hand for each of these structures would be an onerous task. Rather than designing and optimizing parameters manually, we endeavor to create generic descriptions of local symmetry and to utilize machine learning methods, in conjunction with simulation data, to automatically formulate appropriate order parameters for the structures we find. Here we will show that we can use machine learning methods to cluster data into sets of similar structures before the structures have been identified, or to identify systems quickly and efficiently once examples for each structure have been found.

Machine learning (ML) has proven to be a powerful tool in many different fields. Typically, researchers use domain-specific knowledge to create a set of ``descriptors'' which place the data of interest in some high-dimensional space that the ML algorithms can work in. These descriptors should represent the important aspects and invariants of the systems we wish to study. In the field of soft matter, researchers have created ML models using descriptors that are sensitive to particle coordination and local bond angles to identify crystalline phases~\cite{phillips2013,reinhart2017,dietz2017} and glassy solids~\cite{schoenholtz2015} respectively. Reference \cite{keys2011} presents an overview of families of descriptors and order parameters with applications to condensed matter systems. For the study of colloidal self assembly, once we evaluate a set of descriptors for our data, we can apply standard machine learning methods to solve the problems that interest us. In this paper, we present a new method, inspired by Bond Orientational Order (BOO) analysis, to generate descriptors of local particle environments that are sensitive to three-dimensional symmetries. We demonstrate the usefulness of these descriptors by analyzing data from self-assembly of complex structures \emph{via} common machine learning algorithms.

\section*{Bond Orientational Order Analysis}

\begin{figure*}[h!tb]
\begin{center}
\includegraphics[width=\textwidth]{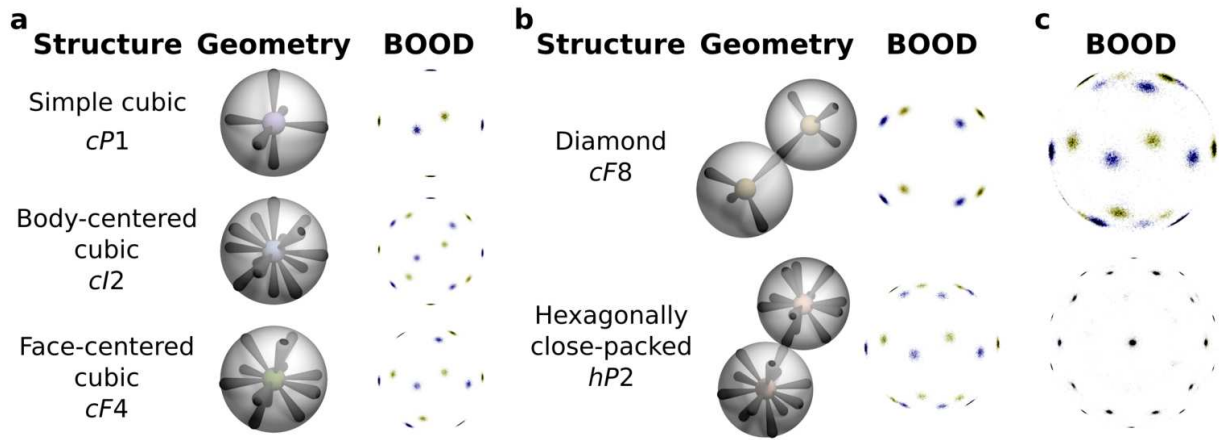}
\caption{Bond orientational order diagrams (BOODs) for various systems. (a) Geometry schematics and BOODs for simple cubic, body-centered cubic, and face-centered cubic structures. (b) Geometry schematics and BOODs of structures with multiple local environment orientations. The BOOD is the superposition of the signals from the bonds of each local environment orientation. (c) BOODs of real face-centered cubic structures with defects. Above, the BOOD appears very similar to that of a hexagonally close-packed structure due to stacking faults. Below, the BOOD appears to exhibit 10-fold symmetry due to polycrystallinity. Points with blue and yellow coronas are only on the front-facing and back-facing side of the sphere, respectively.}
\label{fig:boods}
\end{center}
\end{figure*}

A common method of evaluating the structure of simulated systems is to compare Bond Orientational Order Diagrams (BOODs)\cite{dzugutov1993,roth2000,keys2011,damasceno2012} to those of reference structures. In a BOOD, the bonds --- or vectors drawn between particles, typically within the first neighbor shell --- of all the particles in the system are globally projected in a histogram on the surface of the unit sphere, as shown in Figure~\ref{fig:boods}. Much like a diffraction pattern, the BOOD description of a system can be informative in analyzing the symmetry and quality of a crystal. However, BOOD analysis involves three caveats. First, the presence of multiple crystalline grains can hinder identification of the structure. In the best case this is merely an annoyance, and in the worst case it can lead to the misidentification of a structure. For example, the BOOD of an ABC layered face-centered cubic (FCC) crystal with a stacking fault can appear very similar to the BOOD of an AB layered hexagonally close-packed (HCP) structure, as shown in Figure~\ref{fig:boods}(c). Similarly, FCC structures can also be icosahedrally twinned, which causes the BOOD to exhibit icosahedral symmetry, again leading to structure misidentification. Second, because the orientation of a BOOD is tied to the orientation of the crystal it comes from, point-matching or symmetry detection algorithms\cite{keys2011} would need to be employed to automatically compare to reference BOODs or find high-symmetry axes. Finally, BOODs are graphical metrics and can thus be difficult to quantitatively compare between samples and structures.

In this work we retain the idea of viewing projections of near-neighbor bonds, but rather than arranging them based on the global orientation of the crystal, we orient the bonds of each particle by a local measure: the principal axes of rotation (the eigenvectors of the inertia tensor) of its local neighborhood.
\section*{Local Neighborhood Descriptors}

Global rotational invariance is one of the most basic properties we require of an order parameter. If the order parameters we generate are sensitive to sample orientation, they will be less helpful when identifying the same structure in two different systems, which may have crystallized with two distinct orientations. Ideally, identical bulk structures with different orientations --- or even within the same system in polycrystalline samples --- would be indistinguishable in the values of the order parameters we create. For anisotropic particles in crystals, a good choice of local reference frame could be based on the orientation of the reference particle; however, in plastic crystals, this information would be less useful due to the rotational freedom of the particles.

To achieve global rotational invariance in our algorithm using only local information and without assuming that particles are anisotropic, we orient each particle's local environment based on the principal axes of rotation of its nearest neighbors\footnote{The N nearest neighbors of particle with index $i$ are the N distinct particles with smallest Euclidean distance $||\vec{r_{ij}}||$ from particle $i$, where $i \neq j$.}, represented as point masses. For a given number of neighbors N around particle i, the inertia tensor of the neighborhood is

\[\bar{I}(i,N) = \sum\limits_{j=1}^{N}(\vec{r_{ij}} \cdot \vec{r_{ij}}) \bar{1} - \vec{r_{ij}} \otimes \vec{r_{ij}} \]

\noindent where $\vec{r_{ij}}$  is the vector from particle $i$'s position to the position of its $j$th nearest neighbor, $\bar{1}$  is the identity tensor, and $\otimes$ is the tensor product. We then rotate the points into the principal reference frame for each particle, where the inertia tensor of the neighborhood is diagonal. We accomplish this by finding the eigenvalues $\lambda_i$ and corresponding eigenvectors $\vec{v_i}$ of the inertia tensor. We orient the structure such that the eigenvector with the largest eigenvalue (and moment of inertia) is in the $z$-direction, the second-largest in the $y$-direction, and the smallest in the $x$-direction.

Using the inertia tensor to define the local environment involves three details. First, the result of the diagonalization procedure depends strongly on the number of particles N in the local neighborhood and the symmetry of the structure being studied. For machine learning algorithms, which are often used for very high-dimensional data, we simply concatenate the descriptors computed for several different neighborhood sizes. Second, when the inertia tensor has repeated eigenvalues, diagonalization orients the neighborhood with one (two identical eigenvalues) or two (three identical eigenvalues) remaining degrees of freedom randomly distributed, placing bonds in rings or randomly on the surface of the sphere, respectively. In the latter case both ordered and disordered structures exhibit no distinct intensity peaks, but we emphasize that this only occurs for particular combinations of structure and neighborhood size, so when descriptors are computed for machine learning applications --- using a range of neighborhood sizes --- this ambiguity is not an issue. Finally, in ordered systems with well defined shells of nearest-neighbor particles, there is a degeneracy in terms of which nearest-neighbor particles within the shell the algorithm will find when using numbers of neighbors that do not correspond to full shells. One solution to this problem is to look at the bonds of multiple particles at once, which gives an averaged view of the ways that particles can be placed in neighbor shells. Alternatively, supervised learning methods can naturally learn to map the multiple appearances of a local neighborhood that a single particle may exhibit to a single crystal structure.

Structures can be visualized in the same manner as BOODs\cite{dzugutov1993,roth2000,keys2011,damasceno2012} using histograms of the bonds between neighboring particles, rotated into the reference frame of the local neighborhood as defined above. This procedure forms distinct patterns --- much like BOODs --- for different structures and numbers of neighbors. For several ideal structures with Gaussian noise applied to the positions, we show histograms on the surface of a unit sphere of the four nearest neighbor bonds in this reference frame in Figure~\ref{fig:localBOD} below.

\begin{figure*}[h!tb]
\begin{center}
\includegraphics[width=0.9\textwidth]{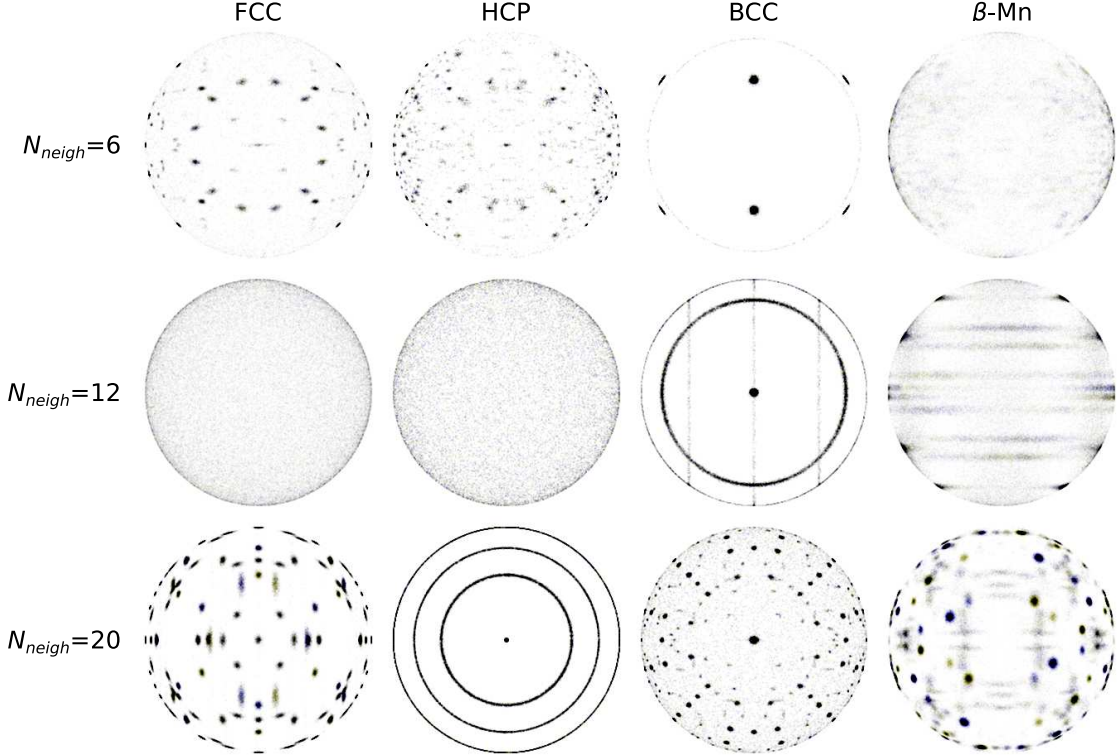}
\caption{Sphere surface histogram of four nearest-neighbor bonds in the local reference frame as defined by the nearest 6, 12, and 20 neighbors for face-centered cubic, hexagonally close-packed, body-centered cubic and $\beta$-manganese structures. FCC and HCP have full neighbor shells at 12 neighbors with diagonal or nearly-diagonal inertia tensors, so they mostly exhibit noise. $\beta$-manganese is a more complex structure with 20 particles per unit cell and exhibits weak patterns at low neighbor counts for this amount of noise.}
\label{fig:localBOD}
\end{center}
\end{figure*}

To create a numerical description of neighboring particle bonds, we use sets of spherical harmonics $Y_l^m(\theta, \phi)$. Because our definition above creates a useful orientation for each particle based on its local environment, we do not have to resort to using rotation-invariant combinations of spherical harmonics~\cite{steinhardt1983} and can evaluate the spherical harmonics for all $l$ and $m$.

We can reduce the spherical harmonics in a number of ways based on the desired application and the capacity of the machine learning methods we plan to use. When classifying individual particles, we use the neighbor-averaged spherical harmonics: for each particle $i$ and a set of spherical harmonics of degree $l$ and order $m$, we define

\begin{equation}
 \label{eq:neighbor_sphs} \bar{Y}_l^m(i, N_n) = \frac{1}{N_n} \left| \sum\limits_{j=1}^{N_n} Y_l^m(\theta_{ij}, \phi_{ij}) \right|
\end{equation}

\noindent where $\theta_{ij}$ and $\phi_{ij}$ are the spherical coordinates of the bond from particle $i$ to particle $j$ in the reference frame of the local neighborhood of particle $i$ as defined above. This averaging method is the same idea that is used when computing $\psi_n$ (and it is identical in the case of $m=\pm l$): the signal from some spherical harmonics would constructively interfere at particular frequencies, while others would exhibit only noise.

We find that neighbor-averaged spherical harmonics work well in a supervised ML setting, but due to the combinatorial degeneracy of placing particles inside neighbor shells, the neighbor-averaged spherical harmonics do not work as well for unsupervised learning algorithms. When training unsupervised models, we can instead look at globally-averaged spherical harmonics; that is, for a particular $l$ and $m$, we generate

\begin{equation}\label{eq:global_sphs}
 \bar{\bar{Y}}_l^m = \frac{1}{N_p N_n} \left| \sum\limits_{i=1}^{N_p} \sum\limits_{j=1}^{N_n} Y_l^m(\theta_{ij}, \phi_{ij})\right|.
\end{equation}

\noindent This is equivalent to taking the spherical harmonic transformation of the local BOODs shown in Figure~\ref{fig:localBOD}. This method sacrifices some of the convenient locality properties of the neighborhood orientation: if there are grain boundaries or defects in the system, they will affect the perceived order of the system as a whole.

\section*{Using Spherical Harmonic Descriptors for Structure Identification}

To validate the usefulness of these descriptors, we study the simulation results of a paper~\cite{engel2015} describing the assembly behavior of a host of complex crystal structures, including clathrates and quasicrystals. We chose this study because it contains some of the most complex crystals in terms of size and structure of the repeat unit that have been predicted so far \emph{via} colloidal and nanoscale self-assembly. The crystals were all obtained using the same two-parameter pair potential, defined as follows:

\[ V(r)= \frac{1}{r^{15}} + \frac{1}{r^3} \text{cos}\left( k(r - 1.25) - \phi \right). \]

\noindent The potential was truncated, shifted, and smoothed to zero at the third maximum to create short-range interaction potentials. The systems were annealed to a low temperature from thermalized initial conditions, creating minimal surface area droplets (not connected through periodic boundary conditions) and columns (connected through periodic boundary conditions in one dimension) of solid. Different combinations of the two independent potential parameters $k$ and $\phi$ produced different crystal structures. Including statistical replicas, this data set contains over 1,100 samples --- a volume that is possible,  but very tedious, for a researcher to analyze by hand over days or weeks. Below we show that by using the spherical harmonics of the neighbor bond distribution --- oriented \emph{via} the local environment --- coupled to standard machine learning methods, we are able to analyze this data set, without \emph{a priori} knowledge of the structures, in an automated manner in under 30 minutes on a common desktop processor. We first pair our descriptors with an unsupervised ML method (clustering \emph{via} Gaussian Mixture Models\cite{dempster1977}) to identify interesting structural regions of phase space and then with a supervised ML algorithm (artificial neural networks) to generate a complete phase diagram from exemplar crystal structures. Detailed descriptions of the analysis performed in all cases are available in the Supplementary Information.

\subsection*{Unsupervised Learning}

After generating the data, analysis of simulation results typically begins by trying to determine which --- if any --- crystal structures are present, with the eventual goal of identifying distinct regions in parameter space where each structure is formed. A simulation dataset could include thousands of combinations of simulation parameters and several replicas for each condition, so being able to lump together similar structures in an automated manner can reduce the required human work by orders of magnitude. This stage of analysis is an ideal application of unsupervised learning, which is often used to group data points together based on some idea of similarity in a high-dimensional space.

We use Gaussian mixture models (GMMs) as implemented in scikit-learn\cite{scikit-learn} to perform unsupervised learning. Briefly, GMMs attempt to create a probability density function that agrees well with the distribution of observed data by using a given number of Gaussian functions in the input space. The number of Gaussian components in the mixture model is typically found by optimizing the Bayesian information criterion (BIC)\cite{schwarz1978}, which measures how well a GMM fits the observed data while penalizing models with many parameters to prevent overfitting.

While GMMs produced by optimizing the BIC usually fit well the density distribution of the dataset they are trained on, the clusters that underlie our data are very commonly not Gaussian-distributed in space. This means that a mapping from the Gaussian component to which a point belongs (or the vector of probabilities for each component) to more meaningful cluster membership is necessary. Several algorithms based on various strategies for generating such a mapping have been proposed over the years\cite{hennig2010,baudry2010,pastore2013,scrucca2016}. In this work we use the method of \cite{baudry2010}, which greedily merges pairs of components based on the largest decrease in Shannon entropy (for observations $i$ and components $j$, $-\sum\limits_{i, j} p_{i,j} ln(p_{i,j})$) caused by merging the pair of components.

Because there are over one thousand simulated systems in the icosahedral quasicrystal dataset from Reference~\cite{engel2015} alone, we use globally-averaged local bond spherical harmonics instead of per-particle local spherical harmonics for our GMMs, as in Equation~\ref{eq:global_sphs}. To find appropriate values for the maximum number of neighbors we use for the local bond descriptors and an appropriate maximum spherical harmonic degree $l$ for the local environment descriptors, we simultaneously optimize these values and the number of Gaussian components in the mixture model using the BIC after taking the 128 principal components of the data if the number of descriptors for each system is greater than 128. In this way we choose a set of descriptors that is most readily fit by Gaussian mixtures with the fewest tunable parameters. In the end, this procedure selects 7 maximum nearest neighbors, a maximum spherical harmonic degree of 7, and a GMM of 15 Gaussian components. In summary, the final set of descriptors for this GMM consist of the 128 principal components of the globally-averaged spherical harmonics for 4 to 7 neighbors with $0 < l \leq 7$ and $0 \leq m \leq l$. We note that the final unsupervised learning results (after merging GMM components) are qualitatively very similar for all combinations of these parameters we tried for less than one close-packed neighbor shell (around 12 neighbors) and with moderate-to-low spherical harmonic degree ($l_{max} \leq 12$).

\begin{figure*}[h!tb]
\begin{center}
\includegraphics[width=0.8\textwidth]{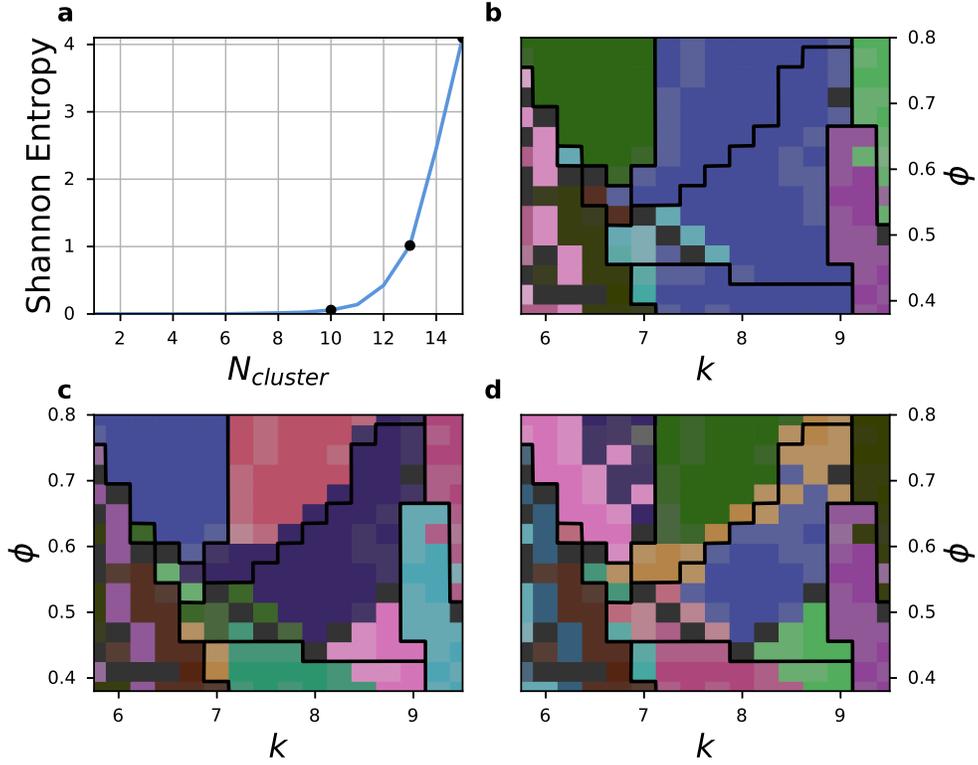}
\caption{Icosahedral quasicrystal dataset phase diagrams generated by unsupervised Gaussian Mixture Models (GMMs). (a) Shannon entropy (blue line) of the quasicrystal dataset as GMM components are successively merged from 15 clusters to one cluster. Merged cluster counts corresponding to (b-d) are indicated by black points. (b-d) Phase diagrams generated by taking the most common predicted cluster type for each parameter point, indicated by the black points in (a). For each selected cluster count, dark gray regions show a poor preference for any single structure among the samples for those parameters. Each type of system as identified by the GMM is assigned a different color. Phase boundaries generated by manual analysis\cite{engel2015} are included for reference as black lines.
}
\label{fig:iqcgmm}
\end{center}
\end{figure*}

After finding a GMM that fits the data well, we can merge Gaussian mixture components to identify the clusters found in our data. Following the method of Baudry\cite{baudry2010}, we sequentially merge pairs of GMM components that yield the largest decrease in Shannon entropy. Figure~\ref{fig:iqcgmm}(a) shows the Shannon entropy as GMM components are merged from the original 15 clusters --- where each cluster corresponds to a GMM component --- down to 1 cluster. In general for this method, the correct number of clusters to use is indicated by an upward elbow in the entropy plot. For data that are not perfectly Gaussian in nature, however, the elbow is smoothed out into more of a curve. Based on this analysis, reasonable choices for the number of clusters to select may be between 9 and 13.

Phase diagrams for three selections of cluster counts, colored by the clusters found at each point in parameter space, are shown in Figure~\ref{fig:iqcgmm}(b-d). Of the structures in this dataset, we find that the models are able to distinguish least clearly between the high-density icosahedral quasicrystal approximant and the disordered region as these are the first components to be merged. In general, we would expect cleaner crystals and crystals with fewer local environments to have more distinct spherical harmonic signatures that are easier for the GMMs to distinguish. Even without knowing how many phases are contained within, the model very accurately maps out the areas associated with the five crystalline regions, the icosahedral/quasicrystal region, and the disordered region. By clustering similar samples together, unsupervised learning can reduce the number of structures in this dataset that must be identified by an expert from over 1,100 to the order of a dozen.

\begin{figure}[h!tb]
\begin{center}
\includegraphics[width=0.49\textwidth]{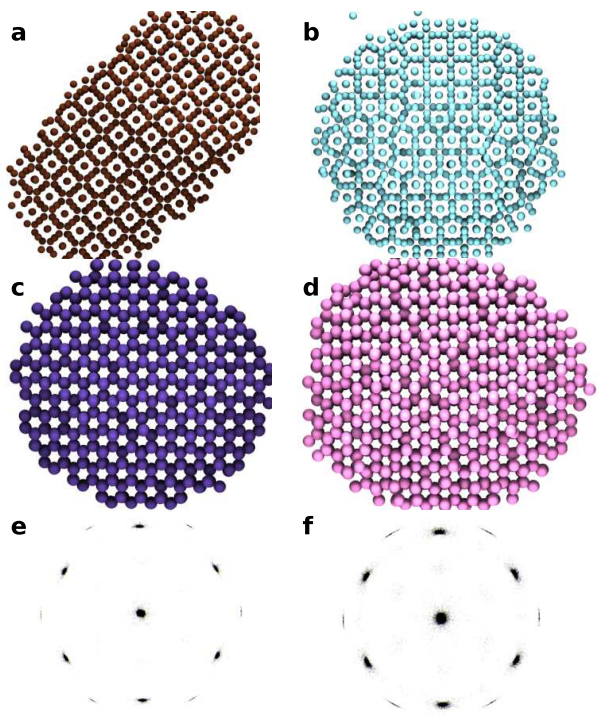}
\caption{Different crystal structures as identified by unsupervised learning of local environments. Colors correspond to the clusters identified by GMM components in Figure~\ref{fig:iqcgmm}(d).  (a-b) Pure $cP8$ and mixed $cP8-tP30$ phases in the $cP8$ region of the phase diagram. (c-d) More-ordered and less-ordered $hP2$ crystals, respectively. (e-f) BOODs of more- and less-well-ordered $hP2$ crystals.}
\label{fig:cp8hp2}
\end{center}
\end{figure}

One interesting observation is the presence of multiple predicted phases in the $cP8$ and $hP2$ regions of the phase diagram. As shown in Figure~\ref{fig:cp8hp2}, on closer inspection we find that one of the $cP8$ region structures of Reference~\cite{engel2015} corresponds to a $cP8$ structure and the others indicate polycrystalline $cP8$ and mixed systems of $cP8$ and $tP30$ (Frank-Kasper $\sigma$) phases. To identify individual particles or crystalline domains as being in local regions consistent with the $cP8$ or $tP30$ phases, we could apply supervised learning to the descriptors of individual particles' local environments instead of globally averaging the descriptors over entire systems. In contrast to the $cP8$ case, the two types of structures found in the $hP2$ region correspond to a more- and less-well ordered version of the same $hP2$ crystal. Because these structures are so similar, it makes sense that they are among the earliest sets of GMM components to be merged. To qualitatively compare the two, we show BOODs of an example of each type of system in Figure~\ref{fig:cp8hp2}(e-f).

\subsection*{Supervised Learning}

\begin{figure}[h!tb]
\begin{center}
\includegraphics[width=0.49\textwidth]{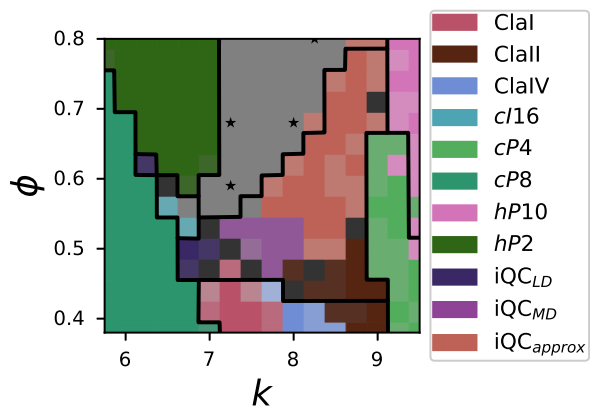}
\caption{Supervised phase diagram generated by a neural network trained on representative structures at particular points in parameter space. Stars indicate locations of training data for the disordered region. Black lines are phase boundaries as identified by hand in~\cite{engel2015}.}
\label{fig:iqcsupervised}
\end{center}
\end{figure}

With supervised learning methods, we can use our local environment descriptors to create order parameters based on our knowledge of which structures are present in the systems we study. We take exemplary simulation data for the five periodic crystal structures, the low- and medium-density icosahedral quasicrystals, the high-density periodic quasicrystal approximant structure, and four points in the disordered region of the phase diagram from the original study~\cite{engel2015} and train a feedforward artificial neural network\footnote{Artificial neural network models were produced with the python library Keras~\cite{chollet2015keras}. } (ANN) to predict the structure (\emph{i.e.}, from which exemplar sample each particle was taken) from the neighbor-averaged spherical harmonics of each particle, as in Equation~\ref{eq:neighbor_sphs}. We use this ANN to construct the phase diagram by finding the most common particle type among all particles in the system at a particular set of conditions. The phase diagram colored by the most prevalent predicted structure in each simulation is shown in Figure~\ref{fig:iqcsupervised}.

Although there are still some $tP30$ samples in the $cP8$ region of the phase diagram, the ANN identifies the whole region as entirely $cP8$ because $tP30$ was not given as a distinct example structure for training. This makes sense because $cP8$ and $tP30$ are similar structures, so the ANN identifies the $tP30$ samples as the nearest structure in descriptor space that it was trained on --- that is, $cP8$. This ability to \emph{generalize} with sensible responses to previously-unseen data is strongly influenced by the choice of descriptors and ML model and cannot be taken for granted, as illustrated by the comparison to Steinhardt order parameters below.

\begin{figure*}[h!tb]
\begin{center}
\includegraphics[width=0.9\textwidth]{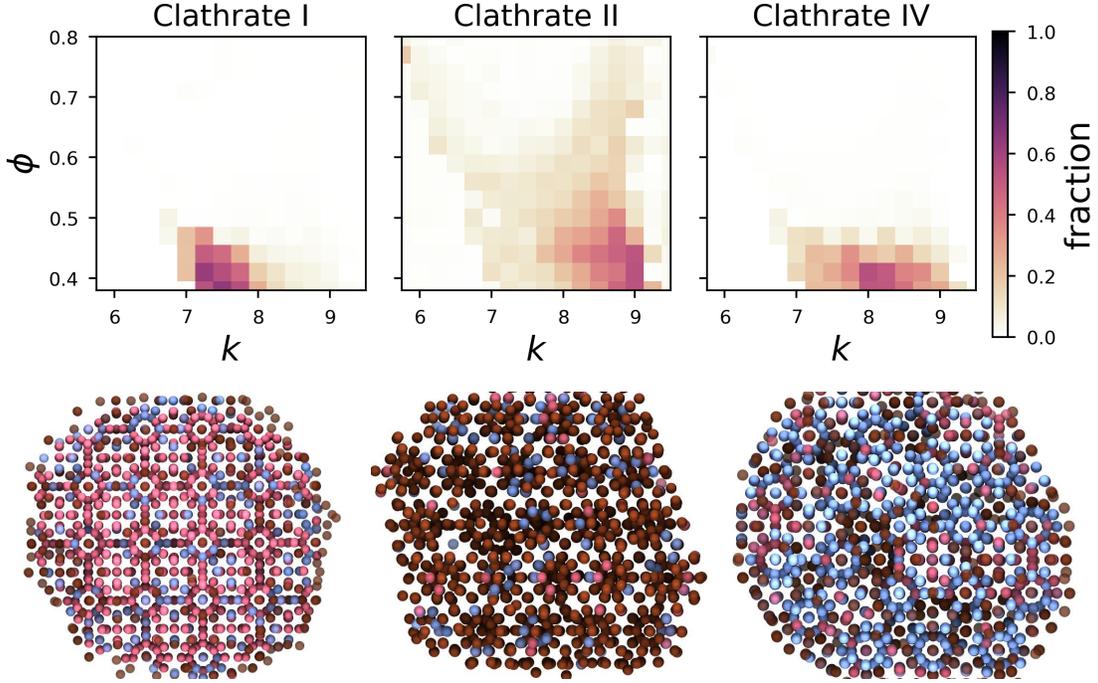}


\caption{Identification of clathrate local environments using supervised learning. Above: Fraction of particles in systems identified as clathrate I, II, and IV, as identified by an ANN. Below: three representative snapshots of simulations with particles colored by their identified structure type. Red: clathrate type I, brown: clathrate type II, blue: clathrate type IV.}
\label{fig:supervisedclathrates}
\end{center}
\end{figure*}

In the original study~\cite{engel2015}, detailed analysis of the clathrate region of the phase diagram was omitted, partly because the clathrates are complicated structures which often appear next to each other in the same simulation to form mixtures. Because the ANN provides a structure estimate for each particle in a system, we can use it to quantitatively identify the prevalence of the three clathrate structures present in this phase diagram, as shown in Figure~\ref{fig:supervisedclathrates}. The ANN finds an abundance of clathrate I at low $k$, II at high $k$, and IV at intermediate $k$, just as was qualitatively described in the original study.

Supervised learning can also be used to help isolate individual grains within a sample, or one structure from another in a mixed system. In Figure~\ref{fig:supervisedclathrates}, we show a few typical systems of clathrates, with each particle colored according to its predicted type based on its local environment. Visually, the ANN is able to distinguish the square tiling arrangement of cage motifs found in clathrate I from the rhombic and triangular arrangement of cage motifs found in clathrates II and IV, even in highly mixed systems.

\subsection*{Comparison to Steinhardt Order Parameters}

We compare our local spherical harmonic descriptors to the Steinhardt order parameters --- which are among the simplest comparable methods and have been extensively used in analysis of 3D ordered systems\cite{steinhardt1983,duijneveldt1992,yan2005,lechner2008} --- to get an idea of their capability. In general, there are many factors that should be carefully considered when comparing two sets of descriptors. Desirable attributes include low computational complexity, high information density, and the ability to be inverted (\emph{i.e.}, to easily compute a structure from a set of descriptor values) or refined (\emph{i.e.}, to be able to generate successively higher-fidelity descriptions). For the purpose of comparing to previous work, here we will focus more concretely on the performance of descriptors in unsupervised and supervised learning applications. The ideal set of descriptors would send sets of observations from each structure into its own compact region in descriptor space. We would also expect the ideal descriptors to place similar structures near each other in this space. Such descriptors would be well-suited for creating informative unsupervised learning models, which typically work using the density of observations or a graph of connectivity in the high-dimensional descriptor space. Because unsupervised learning is generally more difficult than supervised learning, these descriptors would also work well for supervised algorithms.

To qualitatively compare our descriptors to the Steinhardt order parameters we employ t-Distributed Stochastic Neighbor Embedding (t-SNE)\cite{vandermaaten2008,vandermaaten2014}, a nonlinear dimensionality reduction technique. Briefly, t-SNE attempts to create a nonlinear mapping into a lower-dimensional space that preserves point adjacency in a high-dimensional space.

We compare the t-SNE transformation of points from the icosahedral quasicrystal phase diagram for the globally-averaged local descriptors $\bar{\bar{Y}}_l^m$ described in Equation~\ref{eq:global_sphs} (for up to 7 nearest neighbors and spherical harmonic degree 7) and an analogously-defined vector of globally-averaged Steinhardt order parameters $\bar{Q}_l$ (for even $l$ from 2 to 20),

\begin{equation}\label{eq:global_steinhardt}
\bar{Q}_l = \frac{1}{N_p N_n} \sum\limits_{i=1}^{N_p}  \sqrt{\frac{4 \pi}{2l + 1}\sum\limits_{m=-l}^{l}\left| \sum\limits_{j=1}^{N_n(i)} Y_l^m(\theta_{ij}, \phi_{ij}) \right|^2}.
\end{equation}

\begin{figure*}[h!tb]
\begin{center}
\includegraphics[width=\textwidth]{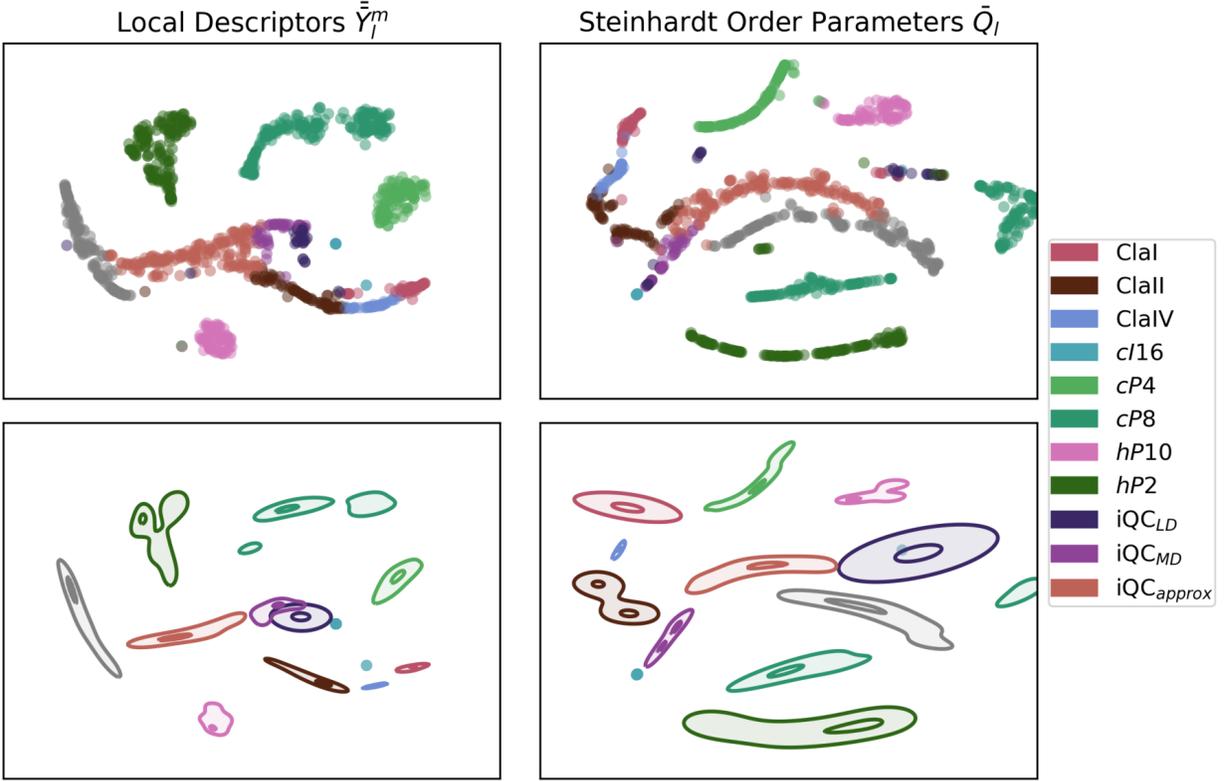}
\caption{t-Distributed Stochastic Neighbor Embedding (t-SNE) distributions of data points generated using globally-averaged local descriptors (left two panels) and a globally-averaged vector of Steinhardt order parameters (right two panels). The plots above show the points after the nonlinear t-SNE transformation directly, while those below show kernel density estimate isosurfaces of the probability distribution. Points for $cI16$ were almost all placed directly on top of one another and the points were plotted rather than using the kernel density estimate in the lower panels.}
\label{fig:tsne_localdescriptors_ql_comparison}
\end{center}
\end{figure*}

\noindent Here we take the neighbors that we sum over as all particles within two standard deviations of the first four nearest-neighbor distances over the whole sample. While this is potentially a somewhat simplistic choice and more complicated methods of using Steinhardt order parameters have been explored\cite{lechner2008}, this is effective to illustrate a baseline comparison of the Steinhardt order parameters and our local descriptors. As shown in Figure~\ref{fig:tsne_localdescriptors_ql_comparison}, both sets of descriptors behave rather similarly under t-SNE dimensionality reduction, with some small differences for the more complex structures. We show the distribution of points in the upper panels and a Gaussian kernel density estimate --- an estimate of the density function using a sum over all points of a Gaussian distance kernel --- of the probability density function for each type of structure in the lower panels (with the exception of the data points for $cI16$, which for both sets of descriptors were nearly all cast to almost the exact same point, impeding display of the kernel density estimate). In both cases, we see that the systems that formed clean, distinct crystals ($cI16$, $cP4$, $hP10$, and $hP2$) are each cast to well-separated regions in the reduced space and that most of the points for the clathrates are adjacent to each other, as we would expect for mixtures of similar structures. The Steinhardt order parameters seem to not perform as well for the $cP8$ phase, which we may guess are split into two regions for the $cP8$-rich and $tP30$-rich systems. Similarly, the Steinhardt order parameters split the low-density icosahedral quasicrystal into multiple parts, but the majority of the low-density quasicrystal points are not adjacent to the intermediate-density quasicrystal. Overall, both sets of descriptors perform well for clean crystals according to this metric, but the Steinhardt descriptors seem to perform somewhat worse for describing the more complex $cP8$, $tP30$, and quasicrystal structures.

While t-SNE dimensionality reduction gives us an idea of how data points are distributed in our descriptor space, its connectivity-based approach does not conclusively show differences in the predictive power of models using these descriptors. To better probe this difference, we generate a phase diagram of the icosahedral quasicrystal dataset using neural networks trained on per-particle Steinhardt order parameters for particle $i$, $Q_l(i)$ (with $l$ from 2 to 20):

\begin{equation}\label{eq:local_steinhardt}
 Q_l(i) = \frac{1}{N_n}\sqrt{\frac{4 \pi}{2l + 1}\sum\limits_{m=-l}^{l}\left| \sum\limits_{j=1}^{N_n(i)} Y_l^m(\theta_{ij}, \phi_{ij}) \right|^2}.
\end{equation}

\begin{figure}[h!tb]
\begin{center}
\includegraphics[width=0.49\textwidth]{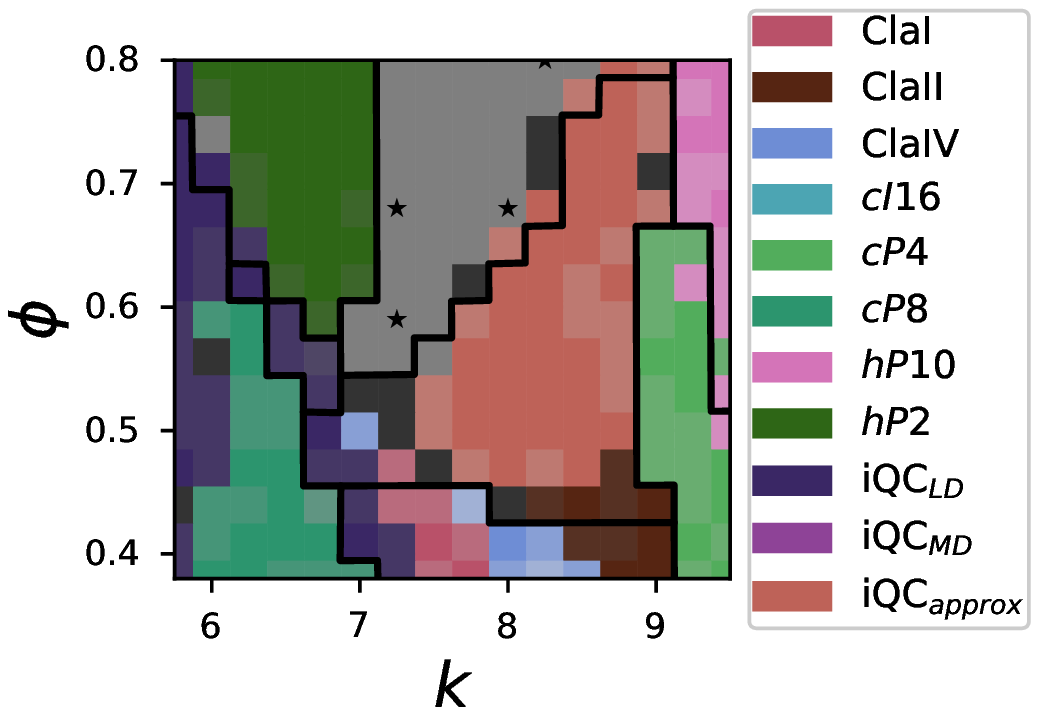}
\caption{Supervised phase diagram generated using a vector of per-particle Steinhardt order parameters $\bar{Q}_l$.}
\label{fig:iqcsupervised_steinhardt}
\end{center}
\end{figure}

\noindent These results should be directly comparable to the phase diagram generated using local descriptors in Figure~\ref{fig:iqcsupervised}. The supervised learning phase diagram generated from Steinhardt order parameter descriptors is shown in Figure~\ref{fig:iqcsupervised_steinhardt}.

We find that, while the phase diagram generated from Steinhardt order parameters agrees with manual analysis in most cases, it differs significantly in how well the ANN model can identify the $cP8$ phase. Looking at these structures manually in more detail, we find that the network trained with Steinhardt order parameters identifies $tP30$, mixed $cP8$-$tP30$, and even pure $cP8$ systems in the high $\phi$ region as the low-density icosahedral quasicrystal structure. Overall, the local environment spherical harmonics seem to generalize better in this case for the purpose of identifying structures than the Steinhardt order parameters.
\section*{Conclusion}

We have introduced a generalized structural descriptor of a particle's local environment that is sensitive to the symmetry of the local neighborhood. These descriptors are scale-free and rotation-invariant, and are useful for supervised, as well as unsupervised, learning of ordered systems. By coupling these numerical descriptions of local environments to common, readily-available machine learning algorithms, we are able to locate interesting structural regions of a complex phase diagram without prior information or to apply our knowledge of the available structures to generate phase diagrams automatically. Because the rate-limiting step of clustering observations into sets of distinct structures happens in an unsupervised manner, this method is highly useful for analyzing results of high-throughput computational experiments. Even though the descriptors are relatively short-ranged, only looking at the 7 nearest neighbors of each particle in this case, they are able to distinguish complicated clathrate structures with dozens of particles in a unit cell --- and even an icosahedral quasicrystal that has no unit cell, but possesses extraordinarily complex orientational order.

In summary, our method allows machine learning algorithms to automatically build order parameters that describe interesting structural behavior from data sets. The machine learning methods and structural descriptors are applicable anywhere that the local environment of a system needs to be characterized, even for complex crystals. We expect our method to be useful in the study of crystal nucleation and growth, glass behavior, and building block design for engineering desirable structures.

\section*{Acknowledgments}
This contribution was identified by Andrew Ferguson (University of Illinois at Urbana-Champaign) as the Best Presentation in the session ``Data Mining and Machine Learning in Molecular Sciences I'' of the 2016 AIChE Annual Meeting in San Francisco. The authors thank Julia Dshemuchadse for helpful discussion of symmetry and structure. M. S. acknowledges support from the University of Michigan Rackham Predoctoral Fellowship program. S. C. G. was partially supported by a Simons Investigator award from the Simons Foundation. Toyota Research Institute (``TRI'') provided funds to assist the authors with their research but this article solely reflects the opinions and conclusions of its authors and not TRI or any other Toyota entity.

\bibliographystyle{unsrt}


\end{document}